\documentclass[runningheads]{llncs}
\usepackage[T1]{fontenc}
\usepackage{graphicx}
\usepackage{color}
\usepackage{multirow}
\usepackage[table]{xcolor}
\usepackage{float}
\usepackage{url}
\usepackage{hyperref}
\usepackage{svg}
\begin{document}
\title{GTR-CTRL: Instrument and Genre Conditioning for Guitar-Focused Music Generation with Transformers}
\titlerunning{Conditioning for Guitar-Focused Music Generation with Transformers}

%
\author{Pedro Sarmento\inst{1}\orcidID{0000-0002-4518-0194} \and
Adarsh Kumar\inst{2} \and
Yu-Hua Chen\inst{3} \and
CJ Carr\inst{4} \and
Zack Zukowski\inst{4} \and
Mathieu Barthet\inst{1}\orcidID{0000-0002-9869-1668}}
\authorrunning{P. Sarmento et al.}
%
\institute{Queen Mary University of London, United Kingdom\\
\email{\{p.p.sarmento,m.barthet\}@qmul.ac.uk} \and
Indian Institute of Technology Kharagpur, India
\and
National Taiwan University, Taiwan \and Dadabots, \url{https://dadabots.com/}\\
}
\maketitle              
\begin{abstract}
Recently, symbolic music generation with deep learning techniques has witnessed steady improvements. Most works on this topic focus on MIDI representations, but less attention has been paid to symbolic music generation using guitar tablatures (tabs) which can be used to encode multiple instruments. Tabs include information on expressive techniques and fingerings for fretted string instruments in addition to rhythm and pitch. In this work, we use the DadaGP dataset for guitar tab music generation, a corpus of over 26k songs in GuitarPro and token formats. We introduce methods to condition a Transformer-XL deep learning model to generate guitar tabs (GTR-CTRL) based on desired instrumentation (inst-CTRL) and genre (genre-CTRL). Special control tokens are appended at the beginning of each song in the training corpus. We assess the performance of the model with and without conditioning. We propose instrument presence metrics to assess the inst-CTRL model's response to a given instrumentation prompt. We trained a BERT model for downstream genre classification and used it to assess the results obtained with the genre-CTRL model. Statistical analyses evidence significant differences between the conditioned and unconditioned models. Overall, results indicate that the GTR-CTRL methods provide more flexibility and control for guitar-focused symbolic music generation than an unconditioned model.



\end{abstract}
\section{Introduction}\label{sec-Intro}


The field of music generation encompasses a diverse set of approaches and has seen progressive improvements towards automation: from musical dice games of the 18th century used by composers to create full compositions \cite{nierhaus2009}, to the earliest instances of computer music making \cite{hiller1979}, and the following explorations of techniques such as rule-based models \cite{Johnson-Laird2002}, Markov Models \cite{Pachet2002}, or genetic algorithms \cite{Papadopoulos98}. Lately, state of the art in machine music creation has been reached with the use of deep learning approaches \cite{Briot2019}\cite{Carnovalini2020}.



This work focuses on automatic music generation using symbolic notations which digitally encode musical score attributes \cite{Raffel2016}. We address the generation of \textit{guitar-focused music}, referring here to music for which the melodic and harmonic arrangements are predominantly conveyed by fretted string instruments, namely guitars and bass guitars. Although fretted instruments are central to tabs, the generation techniques proposed in this work support multiple instruments (e.g. guitar, keyboard, bass, drums). Prior work on symbolic music generation often relies on datasets using formats such as MIDI, MusicXML, and ABC \cite{Dong2020}. We use the DadaGP dataset which, in contrast, is built upon the GuitarPro (GP)\footnote{Available at: \url{https://www.guitar-pro.com/}; alternative software available at: \url{https://sourceforge.net/projects/tuxguitar/}, \url{http://www.power-tab.net/guitar.php}.} format suitable for \textit{guitar-focused symbolic music}. For string instruments, such tab format enables to specify not only \textit{what} music to play but also \textit{how} it should be played \cite{Magnusson2019}. The GP tab format can support expressive renderings of \textit{guitar-focused music} providing information on fingerings and guitar-specific techniques, which are currently not supported by MIDI.


\cite{ShirishKeskar2019} introduced CTRL, a conditional Transformer \cite{Vaswani2017} language model that can control style and content structure for text generation tasks. \textit{Control codes} are special tokens appended to text sequences in order to categorize them and can be used at inference stage as prompts to condition generation features. Inspired by this work, we followed a similar approach by appending special control codes to condition tab generation using a set of instruments and musical genres.

The main contributions of this work are: (1) GTR-CTRL, methods to control a Transformer-based model for automatic tab music generation trained on the DadaGP dataset. The main goal of GTR-CTRL is to provide more flexibility and control at the time of inference. GTR-CTRL is able to be conditioned on instrumentation (inst-CTRL) and musical genre (genre-CTRL); (2) GPBERT, a model  to classify songs into distinct musical genres using the DadaGP dataset, which can be used to assess if the generated music fits expected genres. We hope this work will foster guitar-focused symbolic music and AI research.


\section{Background}\label{sec-RW}

\subsection{Symbolic Music Generation With Deep Learning}

Symbolic music generation techniques with deep learning can be categorized according to the architecture used, namely Variational Autoencoder (VAEs) models \cite{Tan2020}, Generative Adversarial Networks (GANs) \cite{Dong2018}, and models that closely stem from natural language processing (NLP) field, such as Recurrent Neural Networks (RNNs) \cite{Meade2019}, Long Short-Term Memory (LSTMs) \cite{Sturm2016}, or Transformers \cite{Vaswani2017}. The Transformer is a sequence-to-sequence model that is able to learn the dependencies and patterns among elements of a given sequence by incorporating the notion of self-attention. The Music Transformer by Huang et al.\cite{AnnaHuang2019} was first to apply a self-attention mechanism to generate longer sequences of symbolic piano music. Similar seminal works include Musenet \cite{christine_2019}, in which a large-scale Transformer model, GPT-2, was used to generate symbolic multi-instrument music from different musical genres, the Pop Music Transformer \cite{Huang2020}, which uses Transformer-XL \cite{Dai} as a backbone architecture and is able to generate pop piano symbolic music with a better rhythmic structure, and the Compound Word Transformer \cite{Hsiao2021}, that explores novel and more efficient ways  of tokenizing symbolic music for training purposes.

Not much work on guitar-focused symbolic music generation has been done to date despite the proliferation and abundance of tablatures \cite{Macrae2011,barthet2011}. \cite{McVicar2014} presented an automatic guitar solo generator in tablature format, dependent on both input chord and key sequence, by exploring probabilistic models. Regarding guitar tab music generation with deep learning, Chen et al. \cite{Chen2020} presented a fingerstyle guitar generator, trained on a dataset of 333 examples (not using the GuitarPro format), using a Transformer-XL model as a backbone. 

\subsection{Controllable Symbolic Music Generation}

Despite the compelling results of deep learning models for automatic symbolic music generation, difficulties to interpret and control models have persisted. This has fostered research into ways of conditioning and guiding the generation process \cite{Wang2020}. Wang et al. \cite{Wang2020} proposed a VAE model able to generate short piano compositions that could be conditioned on chord structure and style. Similarly, in Music FaderNets \cite{Tan2020}, a VAE architecture is used to generate piano pieces conditioned on rhythm and note density. Regarding genre conditioning, in \cite{Lim2020}, a VAE framework was used to generate MIDI pieces in the style of either Bach chorales or Western folk tunes. Closely related to our work as it used a Transformer architecture \cite{Shih} for conditional symbolic music generation, the Theme Transformer is a novel architecture to generate MIDI scores conditioned on musical \textit{motifs} or themes \cite{Shih}. 

\section{Motivations}\label{sec-motivations}


The main motivation of this work is to devise AI techniques for the production of expressive guitar-focused music which give producers some agency. The adaptation of models proposed in this paper could lead to the development of music making tools for artists/bands, allowing them to stir the generation into different creative directions. To the best of our knowledge, there are no prior guitar-focused symbolic music generation models supporting multiple instruments and that are controllable. The initial experiments we conducted with an \textit{unconditional} model indicate that some degree of control can be achieved through the use of a prompt. For example, by defining one note for each instrument from a desired instrument combination, it is possible to stir the output into a particular instrument arrangement. However, defining said initial notes would often force the key of the composition (which can be a desirable feature or not). By using an excerpt of a song from a given genre as a prompt, the generated music is likely to fit the source musical genre but the content often mimics the initial prompt \textit{motif} which limits creative possibilities. In an attempt to eliminate these constraints and give more creative agency to the user, we investigate alternative token representations facilitating a finer degree of control for tab music generation.



\section{Conditioning Experiments}\label{sec-experiments}

\subsection{DadaGP Dataset}\label{sec-dadagp}

The DadaGP dataset \cite{Sarmento2021} comprises 26,181 songs in two representations: \textit{token format}, which resembles a text representation, and \textit{GuitarPro format}, named after the GuitarPro software for tablature edition and playback. Conversion between these two file formats is possible thanks to a dedicated encoding/decoding tool using PyGuitarPro \cite{PyguitarPro}, a Python library which manipulates GuitarPro files. We used DadaGP as a guitar-focused symbolic music dataset since it represents notes in a syntactic format that is best suited for fretted instruments such as the guitar.

In the DadaGP \textit{token format}, songs start with \verb|artist|, \verb|downtune|, \verb|tempo| and \verb|start| tokens; these \textit{header tokens} are essential for the decoding process from token to GuitarPro format (or vice-versa). Notes from pitched instruments are represented by tokens of the form \verb|instrument:note:string:fret|. This syntax is not only suitable for string instruments since the string/fret combination is eventually mapped to a MIDI note. It can be used for any other pitched instrument supported by DadaGP. Percussive instruments, namely the drumkit, are represented using tokens in the form \verb|drums:note:type|. In order to quantify note duration or rest, the \verb|wait:ticks| token is used. A resolution of 960 ticks per quarter note is used, as in most digital audio workstations (DAWs), and the decoder infers note duration from the tick value. For a more detailed description of DadaGP, please refer to \cite{Sarmento2021}.


\subsection{Model Description}
We used a Transformer-XL model \cite{Dai} as backbone architecture, for it expands on the vanilla Transformer \cite{Vaswani2017} by modifying the positional encoding scheme and introducing the concept of recurrence, enabling it to use information from tokens that occurred before the current segment. To address guitar-focused music generation, we employed the Pop Music Transformer model \cite{Huang2020} which uses a similar architecture to generate piano compositions in MIDI. We trained three variants of a Transformer-XL model configuration, consisting of 12 self-attention layers with 8 multi-attention heads: (i) an \textit{unconditional} model without any possibility of control (apart from side effects of prompting), (ii) a model to control instrumentation (inst-CTRL), (iii) and a model to control musical genre (genre-CTRL). All models were trained for 300 epochs, with a learning rate of $1e-4$ and a batch size of 8 samples. Model parameters were heuristically tuned based on prior experiments.


\subsection{Instrument Conditioning Experiment}

\subsubsection{Instrumentation control tokens:}\label{data-inst}

Following the introduction of control codes presented in \cite{ShirishKeskar2019}, we devised a list of tokens for the instrumentation, i.e. which instruments play in the piece. Before training, instrumentation information is appended at the start of each song as part of header tokens. Tokens for each instrument in a given song are inserted between \verb|inst_start| and \verb|inst_end| tokens. At the time of inference, the model is forced to produce music in the instrumentation given by the tokens. As stated in Section \ref{sec-motivations}, it is possible to control to an extent instrumentation with the \textit{unconditional} model using prompts specifying initial notes for each desired instrument (e.g. \verb|distorted0:note:s6:f0|). However, this indirectly pushes generation towards a specific key, based on the note specified in the prompt (in this example, E minor or major). The proposed inst-CTRL method aims to provide more flexibility.


\subsubsection{Instrumentation inference prompts:}\label{inf-inst}

In order to assess the ability of inst-CTRL to control instrumentation, we devised three distinct initial prompts for the inference stage: \textit{full-prompt}, describing both the instrumentation control codes and an initial note for every selected instrument, \textit{partial-prompt}, which includes instrumentation control codes and a note from one instrument, and \textit{empty-prompt}, containing just the instrumentation control codes. To a varied set of instruments, we selected eight distinct instrument combinations, dividing them into combinations of two, three and four instruments: bass and drums (b-d); distorted guitar and drums (dg-d); distorted guitar, bass and drums (dg-b-d); clean guitar, bass and drums (cg-b-d); distorted guitar, piano and drums (dg-p-d); two distorted guitars, bass and drums (dg-dg-b-d); clean guitar, distorted guitar, bass and drums (cg-dg-b-d); distorted guitar, piano, bass and drums (dg-p-b-d). To compare results from the conditioned and unconditioned models, the \textit{unconditional} case uses a prompt with an initial note for every selected instrument. We generated 1,200 examples for each model/prompt, comprising a total of 150 examples for each of the eight instrument combinations. We defined a limit of 1,024 generated tokens per song using a temperature-controlled stochastic sampling method with \textit{top-k} truncation \cite{ShirishKeskar2019}, as employed in \cite{Huang2020}.



\subsection{Genre Conditioning Experiment}

\subsubsection{Genre control tokens:}
Similarly to the procedure used for instrumentation control (Section \ref{data-inst}), for the genre conditioning experiment we created a list of tokens with musical genre information and placed it at the start of every song in the DadaGP dataset. In order to discard genres with fewer songs, we only selected genres with more than 200 examples in the training corpus. This resulted in around 100 distinct musical genres, with a predominance for genres and subgenres of \textit{Rock} and \textit{Metal}, but also including other genres like \textit{Jazz}, \textit{Folk} and \textit{Classical}.


\subsubsection{Genre inference prompts:}\label{inf-genre}
To assess the genre-CTRL model, we conducted inferences for the following five genres: \textit{Metal}, \textit{Rock}, \textit{Punk}, \textit{Folk}, and \textit{Classical}. Prompts used at inference were chosen similarly to the instrument conditioning experiment. The \textit{full-prompt} provides two measures plus the corresponding genre token. We sampled the first two measures of randomly selected songs from the training corpus in each genre. The \textit{partial-prompt} provides genre tokens, but only uses the first note of every two measure-long snippet. The \textit{empty-prompt} only describes the genre token. The prompt in the \textit{unconditional} case specifies the first two measures but omits the genre token. We generated 20 songs for each of the 5 distinct song prompts, for each of the 5 different genres, resulting in a total of 500 songs. The number of generated tokens and sampling hyper-parameters were kept the same as in the instrument conditioning experiment.

\subsection{Examples}\label{examples}
Following the procedures described in sections \ref{inf-inst} and \ref{inf-genre}, we cherry-picked examples of generated songs for the instrument and genre conditioning models. These examples without any post-processing are made available for listening\footnote{Currently available at: \sloppy\url{https://drive.google.com/drive/folders/1ds5D01YW-8PAkIf-KxbyACOIBEYqk-he?usp=share_link}}.

\section{Evaluation Methods}\label{sec-evaluation}

\subsection{Overall Pitch and Rhythmic Metrics}

In order to assess the overall consistency of the generated results, from both a melodic/harmonic and rhythmic perspective, we utilized two metrics, namely \textbf{pitch class histogram entropy} (PCE) and \textbf{groove consistency} (GC) \cite{Wu2020}. These metrics were computed on examples generated using both the instrument and genre conditioning methods by using the MusPy library, a toolkit for symbolic music generation and evaluation \cite{Dong2020}. We also used metrics specific to each conditioning as outlined in the following sections.

The PCE computes the degree of tonality within a song. Low entropy is obtained when specific pitch classes dominate (e.g. the first and fifth degree of the key), whereas high entropy points towards tonality instability. For inst-CTRL, we computed a combined pitch class histogram entropy figure for every melodic instrument (i.e. drums excluded) in every selected instrument combination and averaged across the different instrument combinations. For genre-CTRL, averaging was done across musical genres. 


The GC as described in \cite{Dong2020}, can be seen as a metric that measures rhythmicity. In compositions where there is a clearly defined rhythm, the grooving patterns between bars remains identical, thus yielding a high score. In opposition, this metric scores low for songs in which there is no rhythmic consistency across measures.

\subsection{Instrumentation Metrics}\label{inst-metrics}

As described in section \ref{inf-inst}, during inference time we prompted the model testing different levels of prompt completeness, in order to investigate the influence of the control codes for instrumentation. To objectively measure the effects, we introduce two new metrics. The \textbf{prompted instrument presence (PIP) score} provides insight on how well the model is capable of generating the instruments that were selected in the control codes and instrumentation prompt. For a given instrument, an empty measure consists of rests throughout the whole duration of the measure. For a given song, the PIP metric indicates the percentage of measures that contain the prompted instruments with respect to all the instruments, as expressed by Equation \ref{eq:pip}:

\begin{equation}
    PIP = \frac{\sum\limits_{i \in P} M_i - E_i}{\sum\limits_{i \in A} M_i}\
    \label{eq:pip}
\end{equation}

where \textit{A} is the set of \textbf{a}ll instruments that appear in that song, \textit{P} is the set of instruments \textbf{p}rompted for a song, hence a subset of \textit{A}, \textit{$M_i$} represents the total number of measures with respect to instrument \textit{i} (including empty measures), and \textit{$E_i$} represents the total number of empty measures assigned to instrument \textit{i}. In order to make the metric more robust, we considered as valid measures, the measures, \textit{$M_i$}, that contain at least one note; we subtracted empty measures, \textit{$E_i$}, to prevent high PIP scores to occur for instrumental parts with many empty measures.

Given that the model may generate parts for instruments that are not in the instrumentation control codes, we implemented an \textbf{unprompted instrument presence (UIP) score}, which measures the percentage of measures with unprompted instruments with respect to the total number of non-empty measures, as expressed by Equation \ref{eq:uip}:

\begin{equation}
    UIP = \frac{\sum\limits_{i \in (A \setminus P)}M_i - E_i}{\sum\limits_{i \in A}M_i - E_i}\
    \label{eq:uip}
\end{equation}

Based on observations from previous experiments, when the model generates only a few notes for a given instrument, the decoder procedures in DadaGP automatically fill the rest of the parts of that instrument with rests, so that the number of measures for that instrument matches the number of measures for the remaining instruments. For example in a 20-measure long composition, if an instrument has only one measure with note information, it will get 19 empty measures assigned to it. As we focus on active musical content (i.e. notes) for unprompted instruments, we discarded empty measures in the UIP formulation.



\subsection{Genre Metrics}\label{genre-metrics}

Aware of the difficulties regarding genre classification, often leading to a lack of consensus even amongst human experts \cite{Oramas2018} and given the sheer volume of generated data, we used a machine-based classifier to assess the genre-CTRL model. Genre recognition is complex considering music in the symbolic domain, which is the domain in which the network outputs music in this work, since acoustic information convey cues for genre identification (e.g. timbre). It is possible to render audio from the generated symbolic notation but the choice and quality of the virtual instruments may affect genre identification. The authors in \cite{Chou2021} introduced MIDIBERT, a Bidirectional Encoder Representations from Transformers (BERT)-based masked language model trained on polyphonic piano MIDI pieces, able to be configured for downstream classification tasks. Following a similar approach, we propose GPBERT, a variant model which was first pre-trained on the DadaGP dataset for a total of 50 epochs, and later fine-tuned for the task of genre classification for 10 epochs. For the latter, we selected a corpus of 800 songs from each of the five genres described in Section \ref{inf-genre}, with a 80/10/10 split between training, validation, and test sets. Achieving an overall test accuracy of 90.67\%, we deemed this model to be suitable for the assessment of the genre-CTRL model. However, this type of assessment present limitations since an AI system is used to classify outputs from another AI system, without human feedback. However, to the credit of the method, the musical genre metadata used for the training of GPBERT were gathered via the Spotify Web API, which provides human-produced genre labels from Spotify's curation teams.

\section{Quantitative Analysis and Discussion}\label{sec-results}


\subsection{Pitch and Rhythm Metrics Results}

\begin{figure}[H]
\centering
\includegraphics[width=1\textwidth]{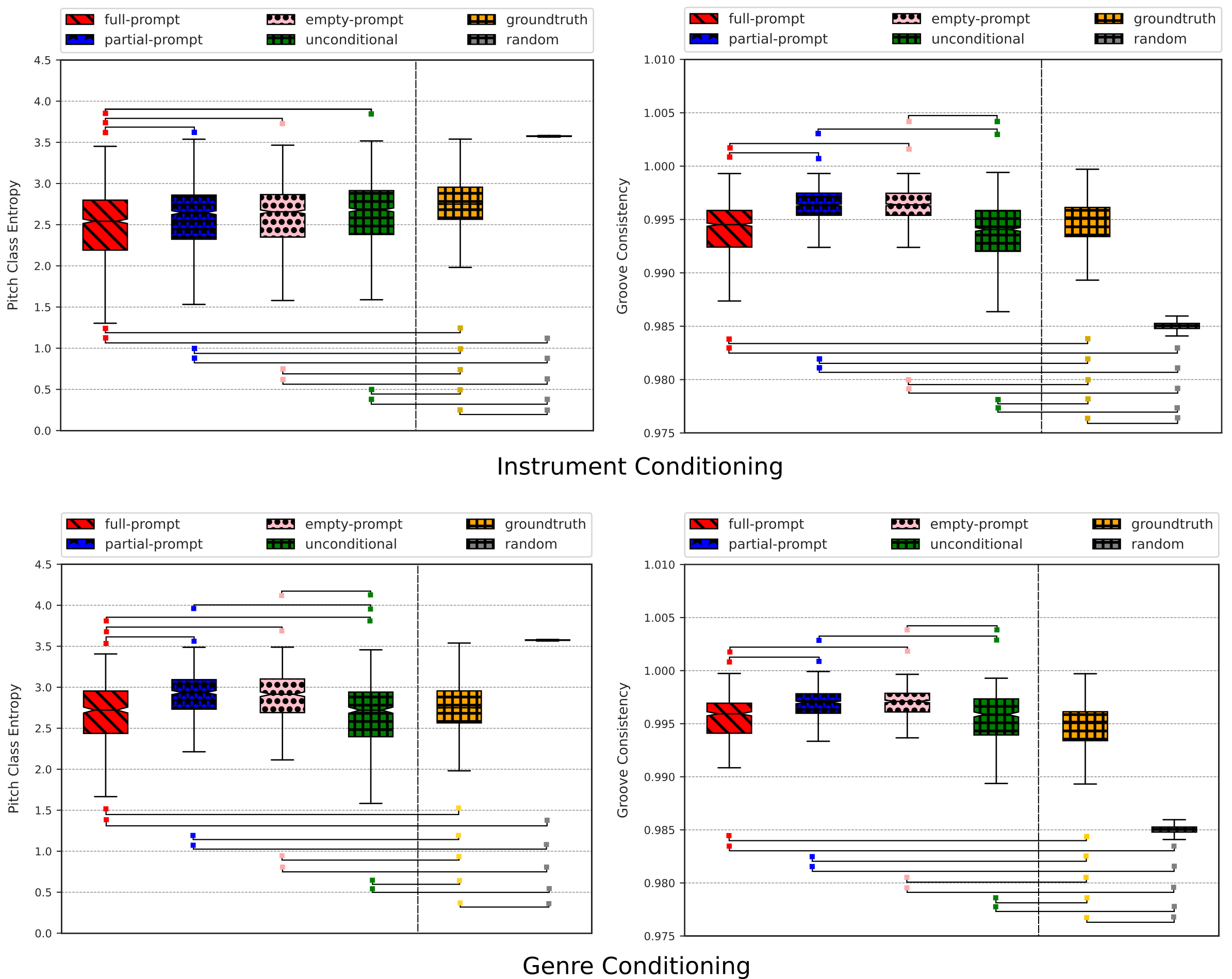}
\caption{Notched boxplots of pitch class entropy (left) and groove consistency (right) for the instrument (top) and genre (bottom) conditioning experiments by comparison with groundtruth and random groups. Brackets indicate significant differences ($p<.001$) in pairwise Wilcoxon signed-rank tests using a Bonferroni-adjusted $\alpha$ level of .008 (.05/6).}
\label{inst-pce-gps}
\end{figure}

We computed the \textbf{pitch class histogram entropy} and \textbf{groove consistency} on the pieces generated for the instrument (1,200 examples in total) and genre (500 examples in total) conditioning experiments. To benchmark the proposed conditioning methods, we used the training corpus as groundtruth data (theoretical best case scenario), and corpi produced by randomizing certain musical attributes (theoretical worst case scenario). For PCE (GC, respectively), the randomized corpus was obtained by randomizing the notes' pitch (rhythm, respectively) in DadaGP songs. We report in this section statistical analyses investigating the effects of the music source (4 different prompts, groundtruth and random) on the PCE and GC dependent variables, with a Type I error $\alpha$ of .05. Notched boxplots of pitch class entropy can be seen on the left plots in Figure \ref{inst-pce-gps} for instrumentation (top) and genre (bottom) conditioning, respectively.

We performed a Kruskal-Wallis rank sum test which yielded a highly significant effect of the music source (4 prompts, groundtruth and random) on pitch class entropy ($H(5)=3169.6$, $p<.001$). Pairwise comparisons were conducted using the Wilcoxon rank sum test and significant differences between conditions are reported using brackets in Figure \ref{inst-pce-gps}. These tests indicate that the three conditional models as well as the unconditional model yield significantly different PCEs compared to the groundtruth and random conditions, for both the instrumentation (top left) and genre conditioning (bottom left).

However, the model-generated examples have PCEs that are much closer to the groundtruth than the random condition. The \textit{full-prompt} obtains PCEs that are significantly different from the other prompts and present the lowest median (the median is slightly below that of the groundtruth). This means that the \textit{full-prompt} condition produces the highest tonal stability on average. However, a low PCE is not a sufficient condition to ensure musical quality given that less tonal variations could be for example due to repetitive structures in the generated music. On average, groundtruth examples comprise $84.6\pm57.0$ measures, whereas generated examples have on average of $34.4\pm24.6$ measures. We hypothesize that the slightly higher tonal stability compared to the training corpus is due to the shorter length of the generated snippets; shorter length often translates into compositions that stay within a given key and without or with only little harmonic modulation which would increase the PCE score. We highlight that it is difficult to precisely define what ``better" results are in terms of PCE due the subjective nature of music. A lower PCE may not necessarily translate into better sounding examples, rather it indicates that the music stays more strongly within a given tonality, on overall. The impact of PCE differences on the musical expectations of listeners should be further assessed in perceptual tests. 






We conducted a Kruskal-Wallis rank sum test to assess the effect of the music source on GC (right plots in Figure \ref{inst-pce-gps}). A highly significant effect can be observed ($H(5)=3666.9$, $p<.001$). For the instrument conditioning experiment (top right), the \textit{partial-prompt} and \textit{empty-prompt} configurations obtained a slightly significantly higher GC median compared to the groundtruth indicating more rhythmic stability. More rhythmic variations tend to occur in longer musical excerpts such as in the groundtruth corpus. The more consistent rhythm obtained for the generated pieces is likely due to the shorter length of the pieces, but it could also be due to the presence of very repetitive structures. Similar conclusions can be made for the genre conditioning experiment.



\subsection{Instrumentation Metrics Results}\label{instrument-cond-results}

For the instrumentation metrics, no comparisons were made with the grountruth and random cases since the metrics are self-explanatory.


\begin{figure}[H]
 \centering
 \includegraphics[width=0.85\textwidth]{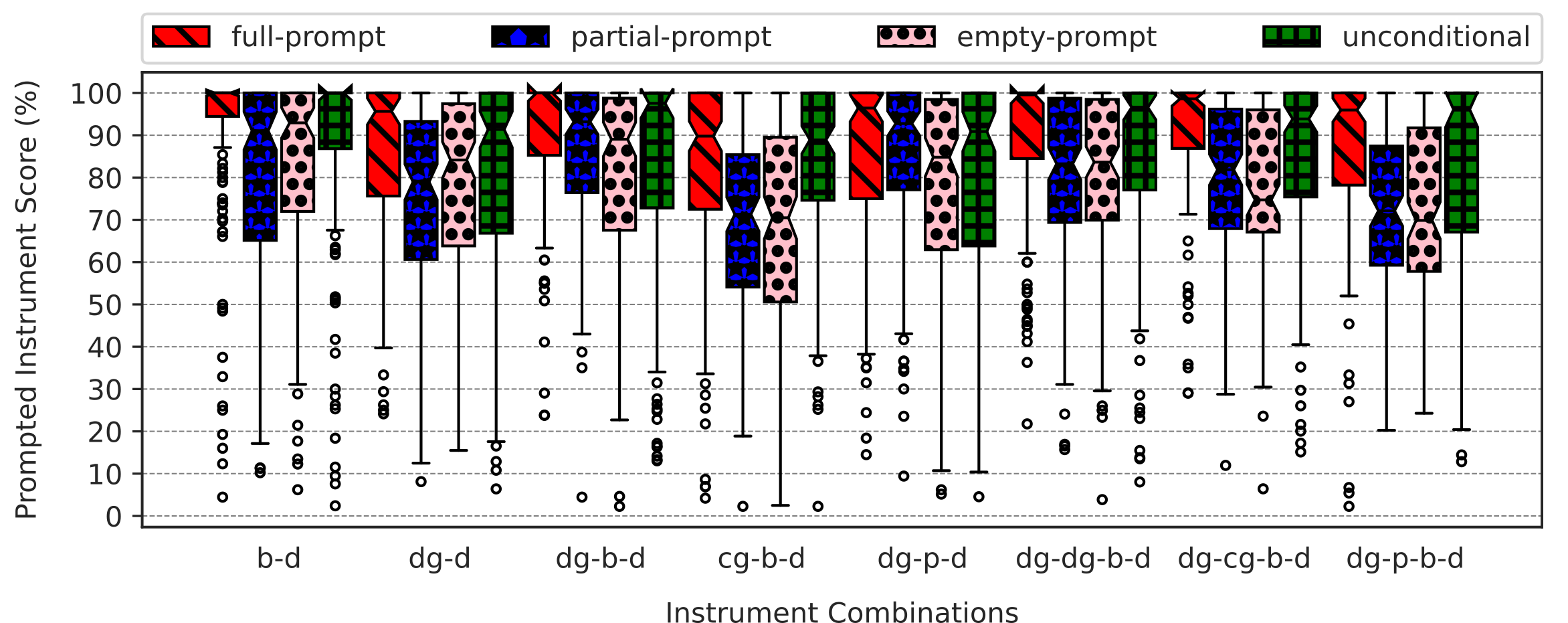}
 \caption{Notched boxplots of the prompted instrument score, for the \textit{full-prompt} (red), \textit{partial-prompt} (blue), \textit{empty-prompt} (pink) and \textit{unconditional} (green) cases, for each instrument combination (higher indicates better performance). }
 \label{fig:prompted-instrument-score}
\end{figure}


Figure \ref{fig:prompted-instrument-score} shows notched boxplots for the \textbf{prompted instrument score}. All the prompt configurations achieved a median PIP score above 70\%. The \textit{full-prompt} achieves the best PIP median (78.2\%, closer to 100\%) for every instrument combination. The \textit{unconditional} outperforms the \textit{partial-prompt} and \textit{empty-prompt} for every instrumentation, with the exception of \textit{dg-p-b-d} (distorted guitar, piano, bass and drums). We believe this is due to a lack of songs with this instrumentation in the dataset yielding less training examples (see instrument distribution in DadaGP in Figure 4(h) \cite{Sarmento2021}). 
A Kruskal-Wallis rank sum test yielded a significant effect of the prompt configuration on the PIP score ($H(3)=304.93$, $p<.001$). Table \ref{tab:pis-upis} reports mean PIP and significant differences occurring between the three conditional models and the unconditional model (Wilcoxon rank sum test with Bonferroni correction). These results show that \textit{full-prompt} obtains significantly better results than the \textit{unconditional}.

\begin{table}
\caption{Mean values of prompted and unprompted instrument score for the \textit{full-prompt}, \textit{partial-prompt}, \textit{empty-prompt} and \textit{unconditional} conditions. Stars indicate significant differences compared to the unconditional model based on pairwise Wilcoxon rank sum test with Bonferroni correction ($p<.001$ for all cases). Best results in \textbf{bold}.}
\centering
\scalebox{0.9}{%
\resizebox{\textwidth}{!}{\begin{tabular}{l|c|c|c|c}
                                & \textbf{Full-prompt} & \textbf{Partial-prompt} & \textbf{Empty-prompt} & \textbf{Unconditional}  \\ 
\hline
\textbf{Prompted Inst. Score}   & \textbf{87.54***}    & 78.23***                & 77.26***              & 83.62               \\ 
\hline
\textbf{Unprompted Inst. Score} & 4.09                 & 4.19                    & \textbf{4.01}         & 5.2                    
\end{tabular}}}
\label{tab:pis-upis}
\end{table}

\begin{figure}[H]
 \centering
 \includegraphics[width=0.8\textwidth]{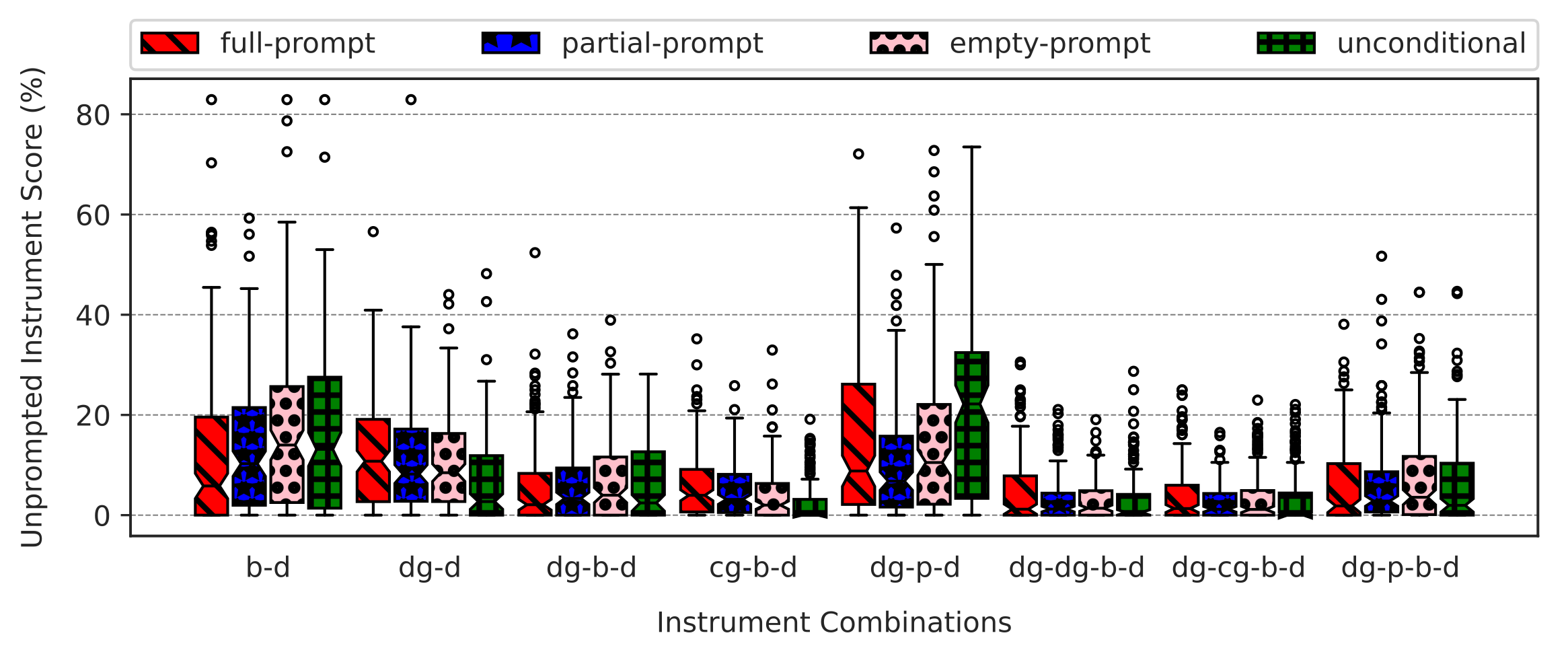}
 \caption{Notched boxplots of the unprompted instrument score, for the \textit{full-prompt} (red), \textit{partial-prompt} (blue), \textit{empty-prompt }(pink) and \textit{unconditional} (green) cases, for each instrument combination (lower indicates better performance). }
 \label{fig:unprompted-instrument-score}
\end{figure}

In Figure \ref{fig:unprompted-instrument-score} we present the notched boxplots for the unprompted instrument score. 

It seems that there is a progressive improvement in avoiding unprompted instruments (lower UIP score) as we move from combinations with less instruments to configurations with more instruments. This may be related to the smaller amount of songs with two or three instruments in DadaGP compared to songs with four or more instruments. However, the \textit{dg-p-d} and \textit{dg-p-b-d} instrumentations obtained worst UIP scores on overall which may be related to the fact that DadaGP only contains four and 25 examples for these configurations, respectively (compared to 4,930 in total for the six remaining selected instrument combinations, the range being [98-3,058]). We conducted a Kruskal-Wallis rank sum test to assess the effect of the instrument combinations on UIP, grouping by number of instruments (respectively 2, 3, and 4 instruments). A significant effect of the number of instruments in instrument combinations can be observed on the UIP score ($H(2)=352.6$, $p<.001$). A pairwise Wilcoxon rank sum test with Bonferroni correction shows significant differences ($p<.001$) between the three instrument combination groups. Combinations with 2, 3 and 4 instruments achieved an average UIP of 6.9\%, 4.35\% and 2.69\%, respectively.





\subsection{Genre Metrics Results}\label{genre-condition-results}


Genre classification scores from GPBERT are shown in Table \ref{tab-genre}, reporting the average softmax score for 100 examples generated for each genre and prompt.
\begin{table}[H]
 \caption{Genre classification softmax scores from GPBERT, for the \textit{full-prompt }(F-P), \textit{partial-prompt} (P-P),\textit{ empty-prompt} (E-P) and \textit{unconditional} (UNC), when conditioned on the genres of \textit{Metal}, \textit{Punk}, \textit{Rock}, \textit{Classical} and \textit{Folk}. Highest results for each genre per prompt in \textbf{bold}.}
\centering
\scalebox{0.8}{%
\begin{tabular}{l|c|c|c|c|c|c}
\multicolumn{1}{l}{}       &     & \multicolumn{5}{c}{\textbf{Genre Classification Score}}                                        \\ 
\cline{3-7}
\multicolumn{1}{l}{}       &     & \textbf{Metal}           & \textbf{Punk}            & \textbf{Rock}            & \textbf{Classical}       & \textbf{Folk}    \\ 
\hline
\multirow{4}{*}{\rotatebox[origin=c]{90}{\textbf{Metal}}}
& F-P & \cellcolor{gray!25}  \textbf{0.6680} &   0.0925          &  0.2028          &  0.0263          &  0.0103  \\
                           & P-P & \cellcolor{gray!25}  0.2217          &  0.1929          &  \textbf{0.4537} &  0.1162          &   0.0155  \\
                           & E-P & \cellcolor{gray!25}  0.2509          &  0.1412          &  \textbf{0.4322} &  0.1577          &  0.0180  \\
                           & UNC   & \cellcolor{gray!25}  \textbf{0.7189} &  0.0675          &  0.1697          &  0.0404          &  0.0035  \\ 
\hline
\multirow{4}{*}{\rotatebox[origin=c]{90}{\textbf{Punk}}}     &  F-P &  0.0041          & \cellcolor{gray!25}  \textbf{0.8126} &  0.1727          &  0.0012          &  0.0094  \\
                           & P-P & 0  .1264          & \cellcolor{gray!25}  0.3406          &  \textbf{0.4807} &  0.0434          &  0.0088  \\
                           & E-P &  0.0997          & \cellcolor{gray!25}  0.2176          &  \textbf{0.5043} &  0.1342          &  0.0441  \\
                           & UNC   &  0.0290          & \cellcolor{gray!25}  \textbf{0.6954} &  0.2464          &  0.0113          &  0.0179  \\ 
\hline
\multirow{4}{*}{\rotatebox[origin=c]{90}{\textbf{Rock}}}      & F-P &  0.0236          &  0.3325          & \cellcolor{gray!25}  \textbf{0.6403} &  0.0021          &  0.0015  \\
                           & P-P &  0.1069          &  0.1661          & \cellcolor{gray!25}  \textbf{0.6401} &  0.0757          &  0.0112  \\
                           & E-P &  0.0863          &  0.0608          & \cellcolor{gray!25}  \textbf{0.5763} &  0.1994          &  0.0772  \\
                           & UNC  &  0.0266          &  0.2207          & \cellcolor{gray!25}  \textbf{0.7487} &  0.0030          &  0.0010  \\ 
\hline
\multirow{4}{*}{\rotatebox[origin=c]{90}{\textbf{Classical}}} & F-P &  0.0338          &  0.0128          &  0.2293          & \cellcolor{gray!25}  \textbf{0.6753} &  0.0488  \\
                           & P-P &  0.0300          &  0.0126          &  0.1883          & \cellcolor{gray!25}  \textbf{0.7035} &  0.0657  \\
                           & E-P &  0.0877          &  0.0644          &  0.3175          & \cellcolor{gray!25}  \textbf{0.4767} &  0.0538  \\
                           & UNC   &  0.0415          &  0.0227          &  0.2638          & \cellcolor{gray!25}  \textbf{0.5940} &  0.0780  \\ 
\hline
\multirow{4}{*}{\rotatebox[origin=c]{90}{\textbf{Folk}}}      & F-P &  0.0692          &  0.0451          &  \textbf{0.4338} &  0.2751          & \cellcolor{gray!25}  0.1768  \\
                           & P-P &  0.0751          &  0.0533          &  0.2904          &  \textbf{0.4323} & \cellcolor{gray!25}  0.1489  \\
                           & E-P &  0.0822          &  0.1141          &  0.3192          &  \textbf{0.3900} & \cellcolor{gray!25}  0.0946  \\
                           & UNC   &  0.0885          &  0.0430          &  \textbf{0.4849} &  0.2901          & \cellcolor{gray!25}  0.0934  \\
\hline

\end{tabular}%
}
   \label{tab-genre}
\end{table}

The \textit{unconditional} and \textit{full-prompt} achieved the highest genre classification results for \textit{Metal}, \textit{Punk} and \textit{Rock}. Whilst the \textit{unconditional} model was able to perform better for \textit{Rock} and \textit{Metal}, the \textit{full-prompt} attains higher scores for \textit{Punk}. It is interesting to note that GPBERT often confuses genres that would also be difficult to discriminate for a human listener (e.g. \textit{Metal}/\textit{Punk} being judged as \textit{Rock}). Overall, the results from the \textit{Folk} genre were often classified as either \textit{Rock} (\textit{full-prompt} and \textit{unconditional}) or \textit{Classical} (\textit{partial-prompt} and \textit{empty-prompt}). For both the \textit{partial-prompt} and \textit{empty-prompt}, when the intended genre is \textit{Metal} or \textit{Punk}, the generated pieces tend to get classified as \textit{Rock}. We believe this is due to the high \textit{Rock}/\textit{Metal} bias in DadaGP, with more than 50\% of tracks in these two genres (see Figure 3 in \cite{Sarmento2021}).

\section{Subjective analysis and Discussion}\label{sec-discussion}

We conducted a subjective analysis of the cherry-picked generated examples. For genre control, particularly in the \textit{empty-prompt} and \textit{partial-prompt} cases, we noted that the model struggles to force a given multi-instrument configuration, frequently producing only one or two instrumental parts. This behavior influences GPBERT genre classification scores which expects multi-instrument compositions for correct classifications. To circumvent this, an interesting approach could be to combine inst-CTRL and genre-CTRL into one single model, accounting for instrumentation and genre controllability.

\begin{figure}[H]
 \centering
 \includegraphics[width=0.85\textwidth]{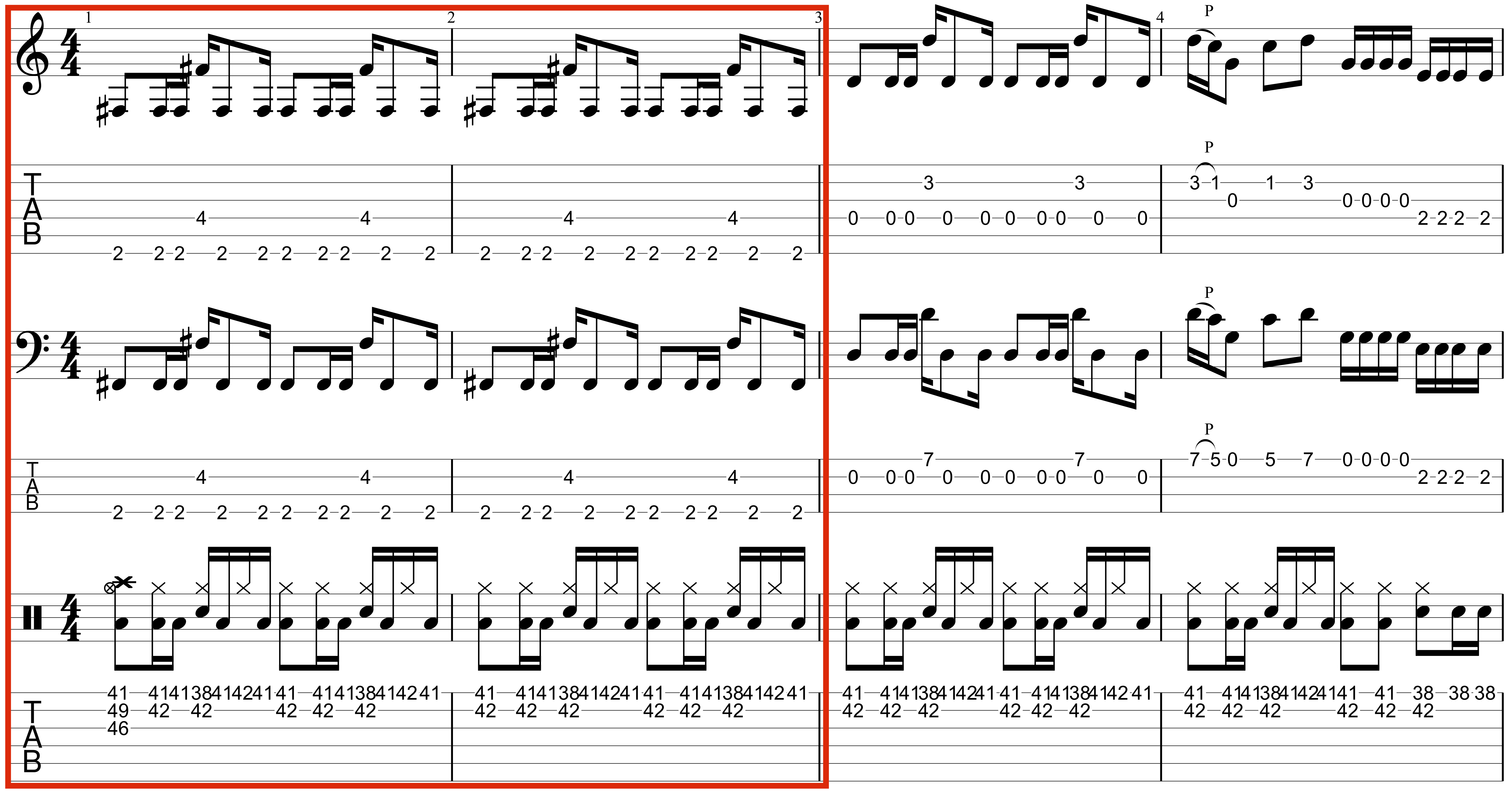}
 \caption{First four measures in \href{https://drive.google.com/file/d/1NhG-ZsUILJstgtYpsZ3bEZIszPqcC_-4/view?usp=sharing}{\underline{id-139}} (prompt from ``Immigrant Song" by Led Zeppelin). Prompt highlighted in red (distorted guitar, bass and drums are visible).}
 \label{fig-led}
\end{figure}

\begin{figure}[H]
 \centering
 \includegraphics[width=0.85\textwidth]{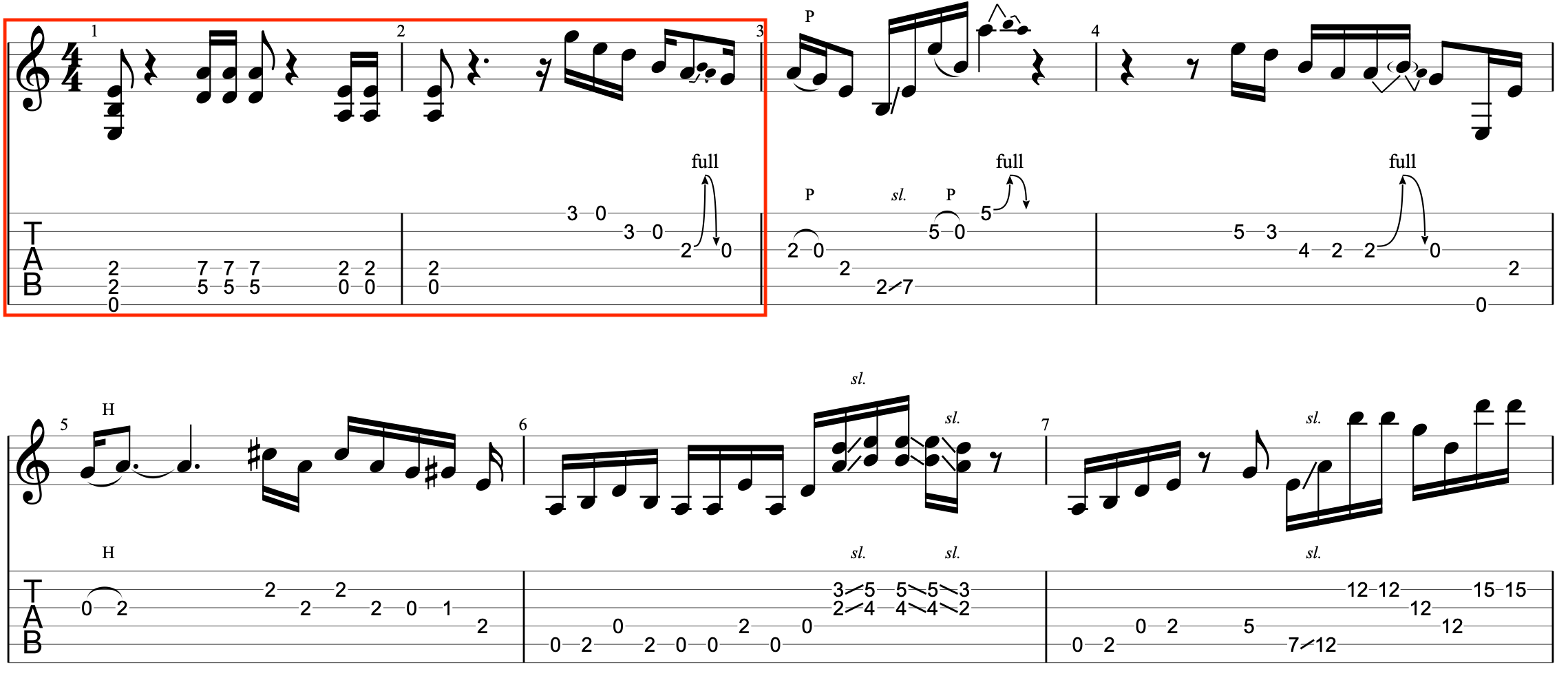}
 \caption{First  measures in \href{https://drive.google.com/file/d/18Clkwoo0Bww7xnJPjaQlKZsTaxyXaUW3/view?usp=sharing}{\underline{id-103}} guitar track (prompt from ``Back In Black" by AC/DC). Measures used as prompt are highlighted in red (only distorted guitar is visible).}
 \label{fig-id}
\end{figure}

Although not a design objective, we found that some of the proposed models with the \textit{full-prompt} genre conditioning could effectively be used as \textit{motif/song continuators}. For example, in song \href{https://drive.google.com/file/d/1NhG-ZsUILJstgtYpsZ3bEZIszPqcC_-4/view?usp=sharing}{\underline{id-139}}\footnote{Underlined song ids are hyperlinked to facilitate listening.}, the model maintains the overall rhythmic structure of the initial \textit{motif}, while adapting it through different chord progressions (see Figure \ref{fig-led}). In \href{https://drive.google.com/file/d/1RJ-vsNl1sPKxj6rrOazKywFHODtfEGbv/view?usp=sharing}{\underline{id-67}} (prompt from the first two measures of ``Sea of Lies" by Symphony X), both guitars perform lead and rhythmic functions, alternating between phrases with guitar solos or chord structures. Furthermore, \href{https://drive.google.com/file/d/1A1ojKqy-8dLw9sUpzu5p2xeMMJR1netY/view?usp=sharing}{\underline{id-459}} (prompt from the first two measures of ``Canon in D" by Johann Pachelbel) stands out for its unusual harmonic exploration that maintains the characteristics of the initial idea, and \href{https://drive.google.com/file/d/18Clkwoo0Bww7xnJPjaQlKZsTaxyXaUW3/view?usp=sharing}{\underline{id-103}} (see Figure \ref{fig-id}) showcases the model's ability to generate novel guitar phrases that fit the key and style of the prompt.



\section{Conclusion and Future Work}\label{sec-conclusion}
In this paper we presented the conditioning of two Transformer-based models, inst-CTRL and genre-CTRL, for the task of guitar-focused symbolic music generation. We studied the performance of these models against an unconditional model, and explored the use of three different prompting strategies at the time of inference. Results show that both models succeed on the respective conditioning tasks: the conditional prompts in inst-CTRL achieve a better performance than the unconditional case in all the metrics, whilst for genre-CTRL, the \textit{full-prompt} outperforms the unconditional model on the \textit{Punk}, \textit{Classical} and \textit{Folk} genres. We believe that the highest scores for the \textit{unconditional} model in \textit{Rock} and \textit{Metal} are due to the large Rock and Metal bias in the dataset, making the \textit{unconditional} model perform well at generating songs in these genres. Despite slightly poorer results, the \textit{empty-prompt} model, by bypassing the need of prompting the model with specific instrument notes that often condition the key/melodic nature of the composition, accounts for an increased level of flexibility, and a completely uninfluenced choice of notes and keys.

In future work, we plan to combine both models to control instrumentation and genre simultaneously. We also intend to explore GPBERT for different downstream tasks, such as music inpainting and artist/composer classification. One interesting research avenue would be to develop a model that is able to generate guitar sections on the style of a particular guitar player, once again using a variant of GPBERT to classify its output, supported by listening tests with expert guitar players. Finally, we envision collaborations with artists/bands willing to engage in human-AI driven compositions.

\section*{Acknowledgments}
This work is supported by the EPSRC UKRI Centre for Doctoral Training in Artificial Intelligence and Music (Grant no. EP/S022694/1).




%
%
%

\bibliographystyle{splncs04}
\bibliography{mybibliography}

\end{document}